# Advanced Signal Processing Techniqes to Study Normal and Epileptic EEG

Debadatta Dash[1]
Dept. of Electrical and Electronics Engineering
Veer Surendra Sai University of Technology (VSSUT)
Burla, India
debadatta2003rta@gmail.com

*Abstract*—**In this paper human normal and epileptic Electroencephalogram signals are analyzed with popular and efficient signal processing techniques like Fourier and Wavelet transform. The delta, theta, alpha, beta and gamma sub bands of EEG are obtained and studied for detection of seizure and epilepsy. The extracted feature is then applied to ANN for classification of the EEG signals.**

## I. INTRODUCTION

In 1875, the English physician, *Richard Canton*, discovered electrical currents in the brain and in 1929 the German psychiatrist, *Hans Berger*, recorded these currents, and named them the Electro-encephalon-gram (EEG) (*Berger, 1929*).The Electroencephalogram is a record of time series of evoked potentials caused by systematic neural activities in a brain. [1]

EPILEPTIC seizures are the result of the transient and unexpected electrical disturbance of the brain. Approximately one in every 100 persons will experience a seizure at some time in their life. [2] .EEG monitoring has an important milestone provide valuable information of those candidates who suffer from epilepsy.[3] Now a days however Magnetic Resonance(MR) Brain Tomography(BT) are used for diagnosis of structural disorders of brain to observe some special illness like epilepsy. The measurement of human EEG signals is done by recording the voltage difference on the scalp by placing electrodes on or inside the brain as specified by internationally recognized 10-20 electrode placement system. [4] There are five major brain waves distinguished by their different frequency ranges. They are: delta ( ) (3Hz and below), theta ( ) (3.5Hz-7.5Hz), alpha ( ) (7.5-13Hz), beta ( ) (14Hz and greater) and Gamma ( ) (31 – 100 Hz) waveforms. [5] Types of epileptic seizures and epileptic syndromes are classified by the Commission of Classification and Terminology of the International League against Epilepsy.

## II. DATA

We used two sets of EEG data; A/Z (Healthy Volunteer, Eyes open) and E/S (Epileptic Subjects during seizure) which is publicly available given by Andrzejak et al [6]. The type of epilepsy was diagnosed as temporal lobe epilepsy with the epileptogenic focus being the hippocampal formation. Each set contains 100 single channel EEG segments of 23.6s duration sampled at 173.61 Hz. [6] Hence each data segments contain 4097 samples collected at the intervals of $1/173.61^{th}$ of 1s. Exemplary (Z038, S001) EEGs are plotted below.

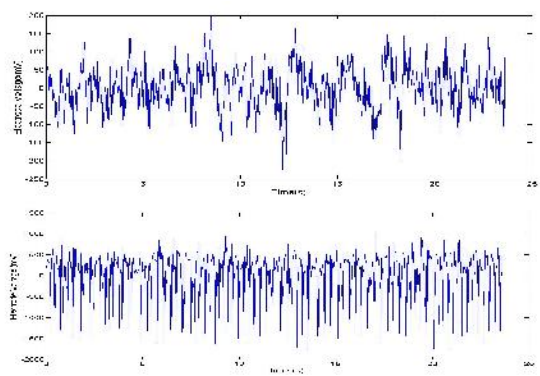

Fig. 1(a). Normal EEG signal, 1(b). Epileptic EEG signal

From Fig. 1. it is seen that the epilectic signal with its higher amplitude and distinct transients is complete different from normal EEG.

## III. FOURIER ANALYSIS

To compare the frequency distribution of both EEG signals we plotted log-log plot of the Periodogram or power vs. frequency plot. Power is explained as the square of the absolute value of Fourier Transform. In each plot so called 1/(f) fluctuation can be seen as the function of frequency.

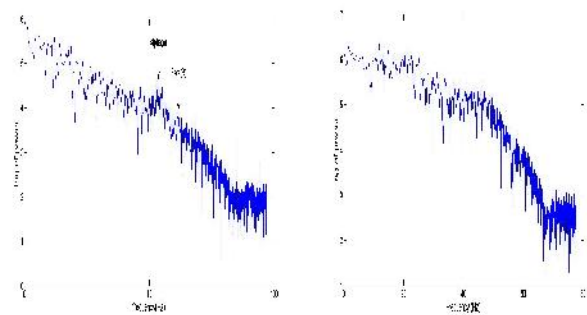

Fig 2. Comparison of log-scale Periodograms of both EEG signals

The peaks that indicate alpha and beta rythms are clearly visible for normal EEG periodogram but not seen in epileptic signal, as the epileptic EEG shows abrupt increase in frequency at different ranges.

## IV. WAVELET ANALYSIS

The basic idea underlying wavelet analysis consists of expressing a signal as a linear combination of a particular set of functions (wavelet transform, WT), obtained by shifting and dilating one single function called a mother wavelet. Since the sampling frequency of the EEG is 173.61 Hz, according to the Nyquist sampling theorem, the maximum useful frequency is half of the sampling frequency or 86.81 Hz. Again to correlate the wavelet decomposition with the frequency changes of the physiological sub bands the EEG band is convolved with a low pass FIR filter and band is limited in 0-60Hz. The basic methodology is as shown below.

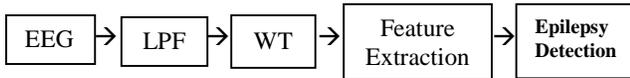

The band limited EEG is decomposed through db-4 wavelet up to $4^{th}$ level and the components retained are $a_4$ (0-4Hz), $d_4$ (4-8Hz), $d_3$ (8-15Hz), $d_2$ (15-30Hz), $d_1$ (30-60Hz). Reconstructions of these five components using the inverse wavelet transform approximately correspond to the five physiological EEG sub bands delta, theta, alpha, beta, and gamma (Fig. 3).

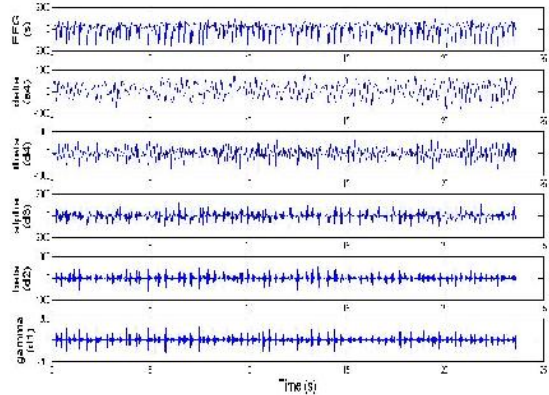

Fig.3. Level 4 decomposition of the band-limited EEG into five EEG sub bands using fourth-order Daubechies wavelet (s=$a_4$ +$d_4$ +$d_3$ +$d_2$ +$d_1$).

Detail parts ($d_1$-$d_4$) shows high frequency components while $a_4$ shows low frequency components. Original signal S is reconstructed by adding all the components as s=$a_4$ +$d_4$ +$d_3$ +$d_2$ +$d_1$. The local low frequency waves concerned with epileptic seizures in the EEG are clearly detected by DWT. The same analysis is done for 100 sets of data of both healthy and epileptic database and the feature is extracted from the delta waves of all the decomposed EEG signals which helped in classification of EEG signals into Epileptic or normal category. The feature extracted are in the category of mean, median, mode, min, max and standard deviation (Std) which showed 97% accuracy for classification of the EEG signals.

The comparison of the features of both normal and epileptic EEG signal of two exemplary data (S001 vs. Z038) is given in the table below.
Table: 1

| EEG Signal | Min | Max | Mean | Median | Mode | Std |
|---|---|---|---|---|---|---|
| Normal | -190 | 185 | 6.816 | 7 | -1 | 478.5 |
| Epileptic | -1765 | 1027 | 47.1 | 187 | 399 | 42.6 |

## V. CONCLUSION

Detecting epileptic seizures from long EEG recordings is a very time consuming and costly task if done visually by a trained professional. The paper proposes a technique for detection and feature extraction of Epilepsy using Fourier, Wavelet Transform with MATLAB. The local low frequency components in each normal and epileptic EEG have been clearly detected by Wavelet Transform but not by Fourier Transform. We conclude that Wavelet Transform is well-suited method for local-time frequency analysis of EEG signals. Once the delta wave is identified and measured, the features like Mean, Max, Min and median was calculated and then ANN is used for the classification with 97% accuracy.